%% file: main.tex
\documentclass[journal]{IEEEtran}
\usepackage{balance}
\usepackage{graphicx}
\usepackage{bm}
\usepackage{amsmath}
\usepackage{amsfonts}
\usepackage{amssymb}
\usepackage{color}
\usepackage{mathrsfs}
\usepackage{upgreek}
\usepackage[utf8]{inputenc}
\usepackage{tikz}
\usetikzlibrary{arrows, calc, decorations.markings, positioning}
\usepackage{cite}
\usepackage{booktabs}
\usepackage{multirow}
\usepackage{amsmath}
\usepackage{algorithm}
\usepackage{algorithmic}
\usepackage{makecell}
\usepackage[flushleft]{threeparttable}
\usepackage{tabularx}
\usepackage{tabularray}         
\usepackage{tikz}               
\usepackage[table]{xcolor}
\usepackage{threeparttable}
\usepackage[caption=false,font=footnotesize]{subfig}
\usepackage[most]{tcolorbox}
\input{SupportDocuments/YFPreamble.tex}

\definecolor{colSignal}{HTML}{FFE5E5}   
\definecolor{colFeature}{HTML}{E5FFE5}  
\definecolor{colDecision}{HTML}{E5F0FF} 
\usepackage{hyperref}					
\hypersetup{
	colorlinks=true,
	citecolor=blue,  
	linkcolor=blue,  
	urlcolor=blue    
}

\title{Movable Antenna-Enabled Multi-Static Cooperative Secure ISAC: Architecture, Mechanisms, and Enabling Technologies}
\author{Yan Yang, Zhendong Li, Jianwei Zhao, Qingqing Wu, Lipeng Zhu, Weimin Jia, and Wen Chen
	\thanks{
		Y. Yang, J. Zhao, and W. Jia are with Rocket Force University of Engineering, Xi’an 710025, China (e-mail: yangyanyy2022@163.com, zhaojianweiep@163.com, jwm602@163.com). 
        Z. Li is with the School of Information and Communication Engineering, Xi’an Jiaotong University, Xi’an 710049, China (email: lizhendong@xjtu.edu.cn).
        Q. Wu and W. Chen are with the Department of Electronic Engineering, Shanghai Jiao Tong University, Shanghai 200240, China (e-mail: qingqingwu@sjtu.edu.cn; wenchen@sjtu.edu.cn).
        L. Zhu is with the School of Interdisciplinary Science, Beijing Institute of Technology, Beijing 100081, China (e-mail: zhulp@bit.edu.cn).
		(Corresponding author: Jianwei Zhao)
	}
	\vspace{-10mm}
}

\begin{document}
\maketitle

\begin{abstract}
The broadcast nature of integrated sensing and communication (ISAC) signals exposes them to physical-layer security (PLS) threats, including eavesdropping and jamming. Conventional fixed-position antennas provide limited spatial flexibility under highly correlated legitimate, sensing, and adversarial channels, while multi-static cooperative ISAC offers distributed spatial diversity but remains vulnerable to information leakage.
To address these limitations, this article proposes a novel movable antenna (MA)-enabled multi-static cooperative secure ISAC framework centered on dual-scale spatial reconfiguration. 
Specifically, continuous MA repositioning reshapes local propagation channels, while distributed multi-static nodes provide complementary network-level spatial diversity. Their synergy transforms local antenna geometry and distributed network geometry into a unified controllable spatial resource for jointly enhancing communication, sensing, and PLS.
This article first presents the system architecture of the proposed framework and its representative application scenarios, followed by an analysis of the physical-layer mechanisms underlying the resulting gains. Subsequently, we elaborate on the key enabling technologies for practical deployment. A case study further demonstrates the communication-sensing-security trade-off and antenna-saving potential of the proposed framework. Finally, open challenges and future research directions are identified toward practical MA-enabled multi-static secure ISAC networks.
\end{abstract}

\vspace{-5.5mm}
\section{Introduction}
\IEEEPARstart{I}{ntegrated} sensing and communication (ISAC) is emerging as an indispensable enabling technology for sixth-generation (6G) wireless networks by allowing communication and sensing functions to share spectrum, hardware, and signal-processing resources.
Emerging applications such as smart cities, the low-altitude economy (LAE), and non-terrestrial networks (NTN) demand wide-area connectivity together with high-resolution and seamless sensing~\cite{10486861}.
However, conventional mono-static and bi-static ISAC systems are constrained by limited coverage and observation perspectives. 
Multi-static cooperative ISAC addresses these limitations by coordinating geographically distributed base stations (BSs) through a central processor (CP) for joint transmission, reception, and information processing~\cite{10273396}, thereby providing network-level spatial diversity and multi-perspective sensing.

The expanded spatial footprint and broadcast nature of multi-static ISAC also increase its exposure to eavesdropping and jamming~\cite{11345706}. Communication signals may be intercepted to obtain confidential data, while sensing waveforms may reveal sensitive target information. Active jammers may further inject jamming signals to degrade both communication and sensing performance.
These threats become more severe when the locations or channel state information (CSI) of adversaries are uncertain, as is common for passive eavesdroppers. Therefore, maintaining communication confidentiality while preserving sensing performance under uncertain and hostile propagation conditions is a key requirement for multi-static cooperative ISAC systems.
Conventional physical-layer security (PLS) techniques with fixed-position antennas (FPAs) rely mainly on beamforming, artificial noise (AN), and power allocation over a predetermined array geometry. Their effectiveness can degrade when legitimate users, sensing targets, and potential eavesdroppers exhibit highly correlated spatial channels, because FPAs provide limited flexibility to reshape unfavorable propagation conditions. Movable antenna (MA) technology introduces an additional spatial control dimension by allowing antenna elements to reposition continuously within a prescribed region in three-dimensional (3D) space. Through position-dependent variations in multipath phase and amplitude, MAs can reshape effective channel gains and correlations~\cite{11386877}. When integrated with multi-static cooperation, such local spatial reconfigurability can be coordinated across distributed nodes, transforming micro-scale channel variations into network-level spatial gains and thereby establishing a dual-scale spatial reconfiguration capability.

Existing studies have established several important building blocks for this direction. MA-enabled ISAC has been investigated through joint antenna position and beamforming design for improving communication and sensing performance~\cite{10839251,10944486}, and extended to networked ISAC systems with distributed transmission and reception~\cite{11204823}. Cooperative multi-static ISAC has been explored through coordinated active and passive sensing~\cite{11203895}. From the security perspective, spatial reconfigurability has also shown considerable potential for improving PLS. For example, absorptive reconfigurable intelligent surfaces (RISs) have been employed to enhance secure transmission in satellite-terrestrial integrated networks~\cite{10700955}, demonstrating the value of propagation reconfiguration under unfavorable legitimate and eavesdropping channel conditions. MA-enabled secure transmission further exploits continuous antenna-position degrees of freedom (DoFs) to reshape legitimate and eavesdropping channels and thereby improve secrecy performance~\cite{11020999}, while joint MA and RIS designs have also been investigated for secure ISAC transmission~\cite{11033708}.
However, these directions have largely evolved independently, and how micro-scale MA-induced channel reconfiguration can be coordinated with macro-scale multi-static diversity for joint communication, sensing, and PLS remains largely unexplored.

Motivated by this gap, this article develops a system-level vision for MA-enabled multi-static cooperative secure ISAC centered on dual-scale spatial reconfiguration. At the micro scale, MA positioning reconfigures local multipath phases, channel gains, and channel correlations; at the macro scale, distributed transceivers provide diverse propagation geometries, observation perspectives, and cooperative spatial DoFs. Their coordination translates local channel controllability into network-level gains for strengthening legitimate links, suppressing information leakage, and maintaining sensing performance. We first establish its network architecture and representative scenarios. We then reveal the underlying physical-layer gain mechanisms. Key enabling technologies are subsequently discussed. A case study illustrates the resulting communication-sensing-security trade-off and antenna-saving potential. Finally, we discuss open challenges and future research directions.

\begin{figure*}[t]
	\centering
	\setlength{\tabcolsep}{0.5pt}   
	\begin{tblr}{
			colspec = {|c|c|c|},            
			hline{1,2} = {dashed, 1.25pt},     
			vline{1,2,3,4} = {dashed, 1.25pt}, 
			row{1} = {bg=cyan!15, abovesep=1pt, belowsep=-3pt},          
		}
		\includegraphics[height=0.215\textwidth]{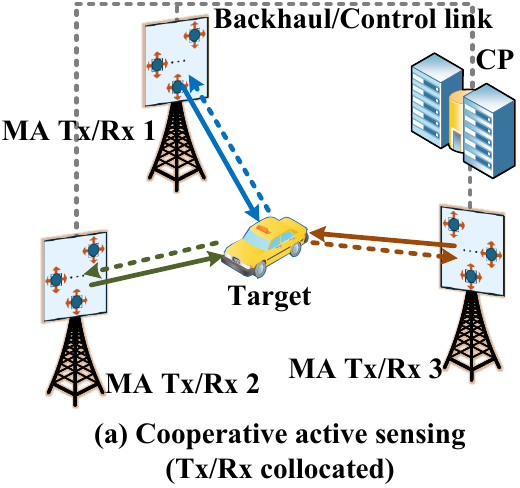} &
		\includegraphics[height=0.215\textwidth]{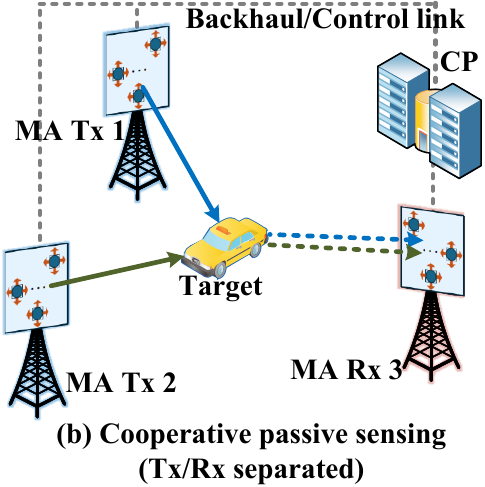} &
		\includegraphics[height=0.215\textwidth]{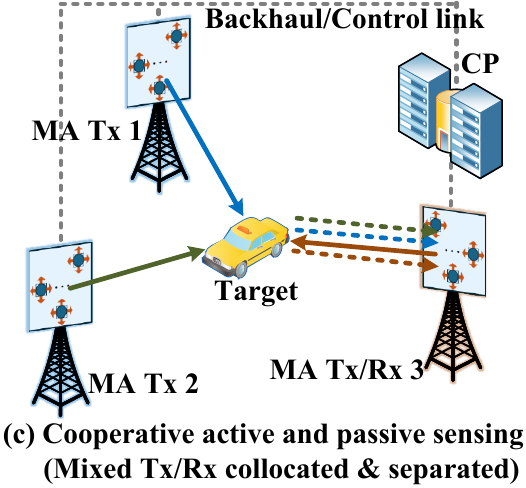} \\
	\end{tblr}
    \vspace{-0.5mm}
	\caption{Types of multi-static cooperative sensing (solid: transmitted signals; dashed: echo signals; Tx: transmitter; Rx: receiver).} 
	\label{multi_BS_sensing}
    \vspace{-4.75mm}
\end{figure*}

\vspace{-3.5mm}
\section{MA-Enabled Multi-Static Secure ISAC Network Architecture}
This section presents the network architecture of the MA-enabled multi-static cooperative secure ISAC framework based on dual-scale spatial reconfiguration. It integrates micro-scale MA reconfigurability with macro-scale multi-static cooperation to jointly support communication, sensing, and PLS.

\begin{figure*}
\vspace{-2mm}
	\centering
	\includegraphics[width=12.250cm]{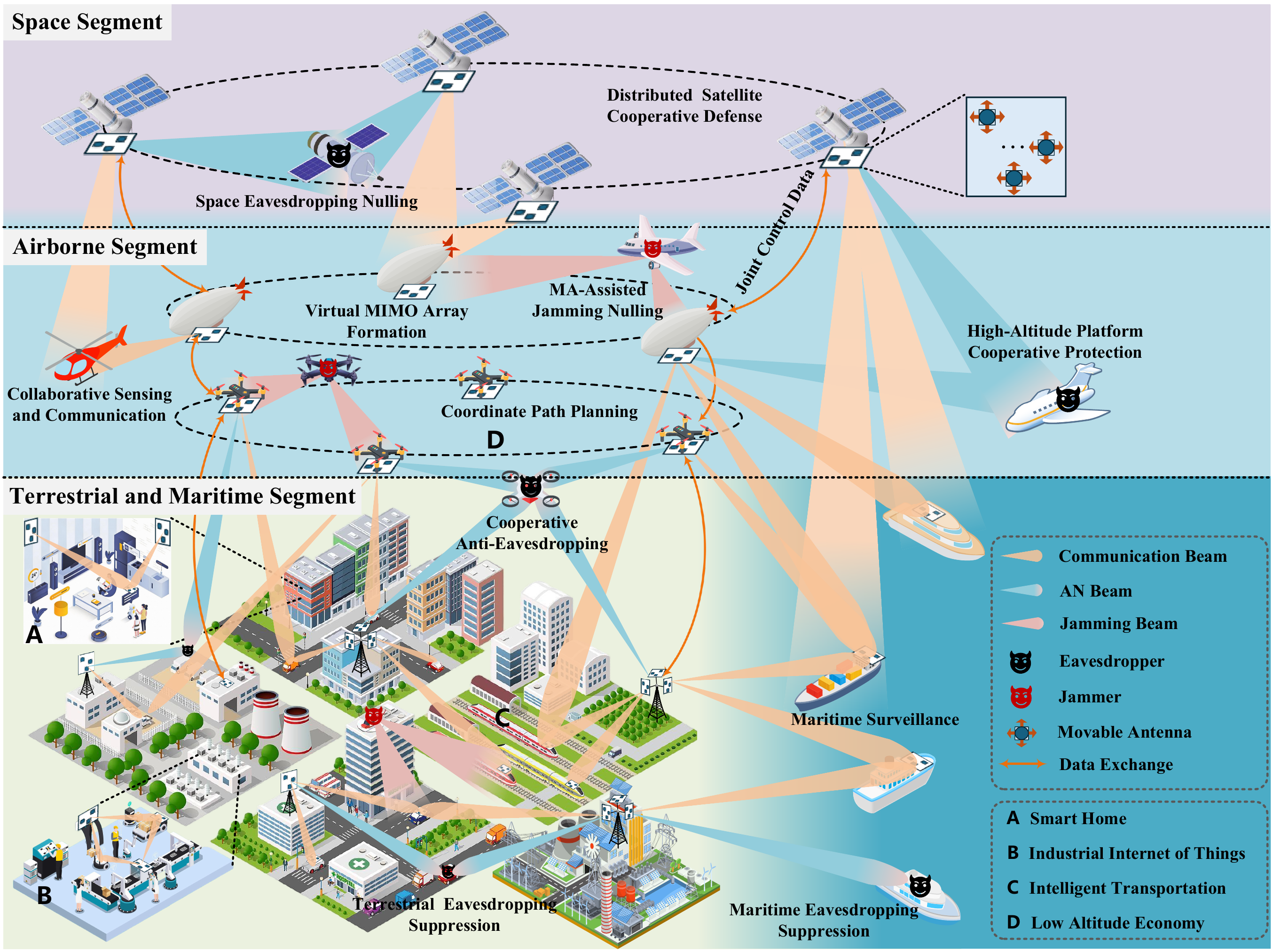}
     \vspace{-0.75mm}
	\caption{Illustration of MA-enabled multi-tier heterogeneous multi-static cooperative secure ISAC networks.}
	\label{fig:magazine_system}
    \vspace{-4.75mm}
\end{figure*}

\vspace{-3mm}
\subsection{Unified Resource Multiplexing}

Communication, sensing, and PLS can share common spectrum, waveform, and hardware resources within the proposed architecture. The spatial DoFs introduced by MAs further enlarge the resource-allocation space, enabling communication beams, sensing waveforms, and AN to be jointly coordinated~\cite{9755276}. In particular, communication signals can be designed to simultaneously support data transmission and target sensing, while appropriately designed AN can suppress information leakage and contribute additional sensing energy. Such coordinated resource multiplexing improves the utilization of limited spectral and spatial resources and enables a flexible balance among communication quality, sensing performance, and security requirements.

\vspace{-3.5mm}
\subsection{Micro-Scale MA Hardware Architecture}

At the micro scale, cooperative BSs can employ MAs whose positions are continuously adjustable within predefined local regions. The antenna elements can be connected to radio-frequency chains through flexible feeding structures and repositioned using rail-, motor-, or other electromechanical mechanisms, introducing antenna position as an additional controllable dimension. By adjusting antenna positions, each BS can modify the phases and amplitudes of multipath components, thereby reshaping local channel gains and spatial correlations. This fine-grained reconfiguration can strengthen desired links, reduce signal leakage toward potential eavesdroppers, and adapt sensing responses to surrounding targets.

\vspace{-3.5mm}
\subsection{Macro-Scale Multi-Static Cooperative ISAC}

Multi-static cooperation extends sensing coverage and observation diversity beyond conventional mono-static configurations. According to the transmitter-receiver relationships among distributed BSs, multi-static cooperative sensing can be categorized into three modes~\cite{10273396}, as illustrated in Fig.~\ref{multi_BS_sensing}.
\begin{itemize}
    \item \textbf{Cooperative Active Sensing:} Participating BSs actively transmit dedicated ISAC signals or reuse communication waveforms, while collocated transmit-receive nodes cooperatively process the target echoes. Multiple active sensing links provide diverse observation perspectives and support high-accuracy localization and parameter estimation. This mode is particularly suitable for high-precision applications such as vehicle-to-everything (V2X) and intelligent transportation. This mode incurs additional transmission and coordination overhead and, when simultaneous transmit-and-receive operation is adopted, requires effective self-interference suppression.
    \item \textbf{Cooperative Passive Sensing:} Receiving BSs exploit signals emitted by other cooperating nodes rather than transmitting dedicated sensing signals. This reduces additional sensing-transmission overhead and improves the operational covertness of passive receivers, making the mode attractive for covert monitoring and spectrum surveillance. Its performance depends on transmitter-receiver geometry, waveform availability, and the synchronization required by the adopted processing strategy.
    \item \textbf{Cooperative Active and Passive Sensing:} Active and passive sensing links coexist within the same network, enabling flexible transmit-receive role assignment and joint processing of echoes from local and cooperative transmissions. This mode offers a practical trade-off among sensing accuracy, coverage, hardware overhead, synchronization requirements, and operational covertness, and is well suited to heterogeneous air-ground and low-altitude networks.
\end{itemize}
These modes enable distributed sensing roles and network topologies to be configured according to coverage, fronthaul, synchronization, and hardware constraints, forming a macro-scale spatial architecture complementary to MA-enabled local channel reconfiguration.

\vspace{-3mm}
\subsection{MA-Enabled Multi-Tier Heterogeneous Network}

As illustrated in Fig.~\ref{fig:magazine_system}, the proposed framework can be extended to space-air-ground-sea integrated networks, where heterogeneous terrestrial and non-terrestrial nodes jointly provide complementary coverage and sensing perspectives.
\begin{itemize}
    \item \textbf{Space Segment:}
    Satellite-based NTNs are becoming an important component of 6G ubiquitous connectivity, with direct-to-smartphone transmission complementing terrestrial coverage~\cite{10486861}. Multi-satellite constellations provide wide-area coverage together with distributed transmission and observation perspectives, which can support cooperative sensing and secure transmission. Their largely predictable orbital trajectories facilitate long-term coordination, while satellite-mounted MAs can further adjust local propagation conditions to strengthen legitimate links and reduce signal exposure toward potential eavesdroppers.
    \item \textbf{Airborne Segment:}
    Unmanned aerial vehicles (UAVs) and high-altitude platforms offer flexible and rapidly changing observation geometries for communication relaying and cooperative sensing. Coordinated trajectory planning controls the large-scale spatial relationships among airborne nodes, while MA positioning provides fine-grained local channel adaptation. Their joint use improves spatial flexibility for confidential transmission and sensing in dynamic low-altitude environments.
    \item \textbf{Terrestrial and Maritime Segment:}
    Terrestrial BSs and maritime nodes often operate under complex propagation conditions caused by blockage, multipath, sea-surface reflections, and mobility. Multi-static deployment provides complementary communication and sensing links, while MA positioning exploits favorable local propagation conditions. Coordinated beamforming and antenna-position adjustment can strengthen desired links while reducing signal leakage and interference in dense urban and maritime environments.
\end{itemize}

\vspace{-5.5mm}
\subsection{Application Scenarios}

The dual-scale spatial controllability of the proposed architecture can support several representative 6G applications, as illustrated in Fig.~\ref{fig:magazine_system}.
\begin{itemize}
    \item \textbf{Smart Home:}
    Indoor environments are typically characterized by rich multipath propagation and highly location-dependent channel conditions. MA-equipped WiFi sensing nodes can adjust antenna positions to exploit favorable multipath components for simultaneous communication and contactless sensing, while security-oriented beamforming and spatial nulling reduce signal leakage toward unauthorized devices.
    \item \textbf{Industrial Internet of Things:}
    Industrial environments contain dense metallic structures, moving machinery, and strong reflections, resulting in complex and time-varying propagation conditions. Multi-static sensing provides complementary observations for tracking industrial devices, while MAs exploit local multipath variations to improve channel separability, supporting low-latency control communication and resilience against eavesdropping and wireless jamming.
    \item \textbf{Intelligent Transportation:}
    Roadside BSs and distributed nodes can observe vehicles and surrounding traffic from multiple perspectives, improving sensing coverage in obstructed or non-line-of-sight regions. MA further adapts communication beams and AN to changing vehicle positions, helping preserve legitimate-link quality while reducing information leakage.
    \item \textbf{LAE:}
    Low-altitude networks involve highly mobile UAVs and rapidly varying propagation conditions. Multi-static cooperation provides complementary observations for trajectory monitoring and airspace awareness, while coordinated MA positioning, beamforming, and AN transmission can enhance the confidentiality and anti-jamming resilience of UAV communication links.
\end{itemize}

\begin{figure*}[t]
	\centering
	\setlength{\tabcolsep}{0.2pt}   
	\begin{tblr}{
			colspec = {Q[c,0.25\textwidth] Q[c,0.285\textwidth] Q[c,0.285\textwidth]}, 
			hline{1} = {dashed, 1pt},                    
			hline{2} = {2-3}{dashed, 1pt},               
			hline{3} = {dashed, 1pt},                    
			vline{1,2,3,4} = {dashed, 1pt},              
			row{1,2} = {bg=cyan!15},                     
			colsep = 0pt,                                
			rowsep = 0pt,                                
		}
		\SetCell[r=2]{c}                             
		\begin{minipage}{\linewidth}
			\vspace{-1.5pt}
			\hspace{0pt}\scriptsize (a)
			\vspace{1.5pt}
			\par
			\centering
			\includegraphics[width=0.95\linewidth]{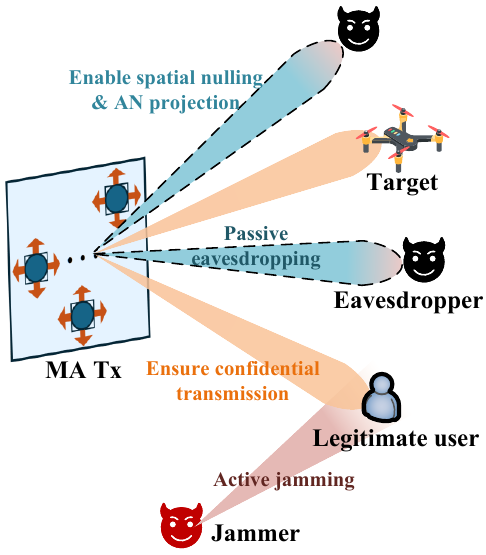}
			\vspace{0.5pt}
		\end{minipage}\label{MA_com_sen_secure_model}
		&
		\begin{minipage}{\linewidth}
			\vspace{0.5pt}
			\hspace{0pt}\scriptsize (b)
			\vspace{0.5pt}
			\par
			\centering
			\includegraphics[width=0.875\linewidth]{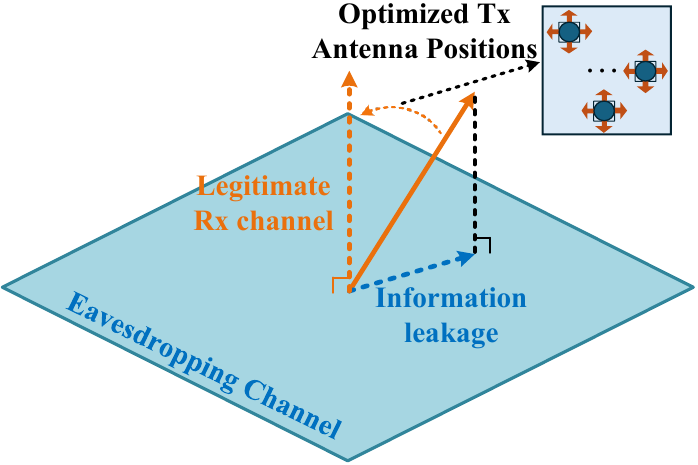}
		\end{minipage}\label{MA_LU_Mechanism}
		&
		\begin{minipage}{\linewidth}
			\vspace{0.5pt}
			\hspace{0pt}\scriptsize (c)
			\vspace{0.5pt}
			\par
			\centering
			\includegraphics[width=0.875\linewidth]{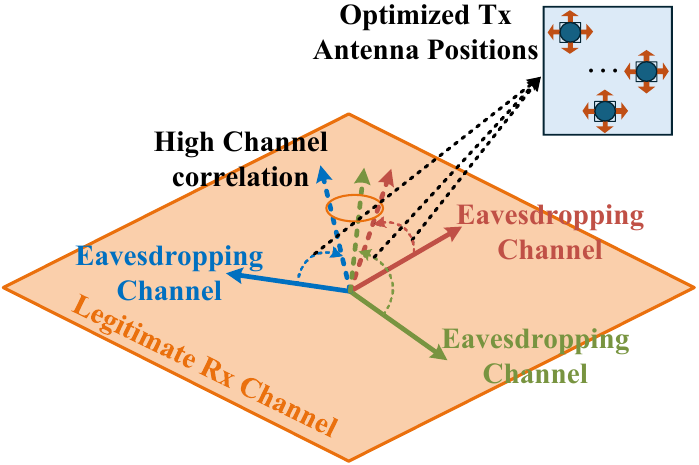}
		\end{minipage}\label{MA_EVE_Mechanism}
		\\
		&
		\begin{minipage}{\linewidth}
			\vspace{0.5pt}
			\hspace{0pt}\scriptsize (d)
			\vspace{0.5pt}
			\par
			\centering
			\includegraphics[width=0.875\linewidth]{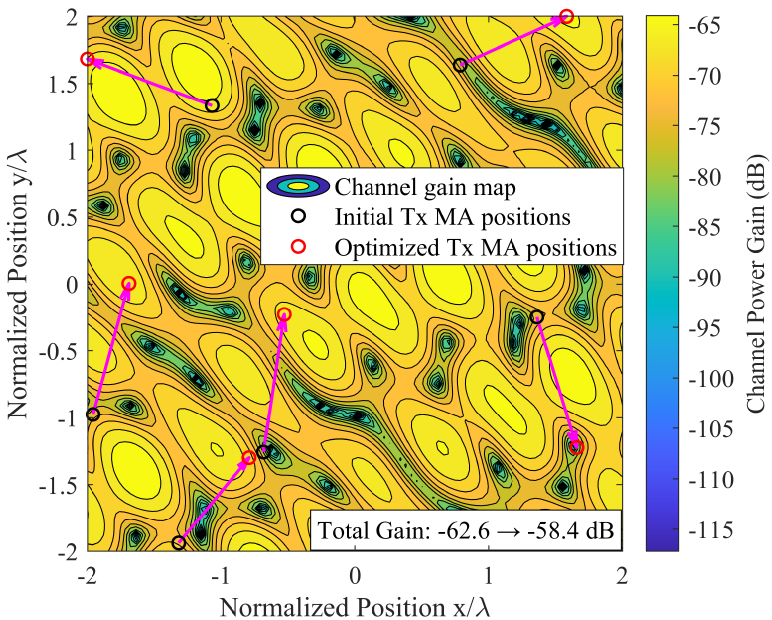}
            \label{MA_DL_map}
		\end{minipage}
		&
		\begin{minipage}{\linewidth}
			\vspace{0.5pt}
			\hspace{0pt}\scriptsize (e)
			\vspace{0.5pt}
			\par
			\centering
			\includegraphics[width=0.875\linewidth]{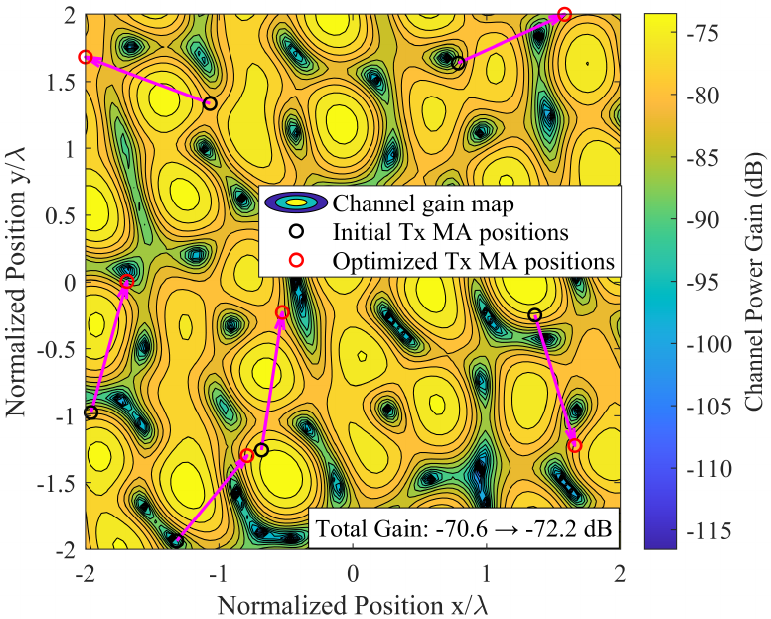}
            \label{MA_EVE_map}
		\end{minipage}
		\\
	\end{tblr}
	\caption{Illustration of channel correlation reconfiguration mechanism in MA-enabled secure ISAC: (a) System model;  (b) Reducing channel correlation between the legitimate Rx and the eavesdropper; (c) Improving channel correlation among eavesdroppers; (d) Channel power gains between the transmit MA of the Tx and the single FPA of the legitimate Rx; (e) Channel power gains between the transmit MA of the Tx and the single FPA of the eavesdropper.}
	\label{MA_Mechanism}
    \vspace{-4.750mm}
\end{figure*}

\vspace{-3.15mm}
\section{Physical-Layer Gain Mechanisms}\label{Physical-Layer Mechanisms}
This section explains the physical-layer mechanisms underlying the proposed framework, focusing on the following three complementary gain mechanisms.

\vspace{-3mm}
\subsection{Proactive Channel Reconfiguration via MA}
Conventional FPAs can only optimize beamforming over a predetermined spatial configuration. When legitimate users, sensing targets, and eavesdroppers exhibit highly correlated angular or multipath characteristics, beamforming-based PLS techniques may suffer from limited spatial separability and degraded secrecy performance. In contrast, MAs introduce an additional spatial control dimension by continuously adjusting antenna positions within a prescribed region. Since antenna displacement changes the relative phases of multipath components, MA positioning can proactively reshape the effective channel gains and correlations rather than merely adapting to a fixed propagation environment~\cite{11345706}. By coordinating MA positions across multiple BSs, the proposed framework can promote constructive signal combination toward legitimate users and sensing targets while inducing destructive interference toward potential eavesdroppers, thereby enhancing physical-layer confidentiality.
The underlying mechanism is illustrated in Fig.~\ref{MA_Mechanism}. Fig.~\ref{MA_Mechanism}(a) depicts a typical MA-enabled secure ISAC scenario. 
As shown in Fig.~\ref{MA_Mechanism}(b), MA position optimization reduces the correlation between legitimate and eavesdropping channels, increasing their spatial separation and facilitating beamforming and AN suppression. For multiple eavesdroppers, Fig.~\ref{MA_Mechanism}(c) shows that MA positioning can instead increase the correlation among leakage channels, concentrating them into a smaller spatial subspace so that limited AN and nulling resources can suppress them more efficiently while preserving spatial resources for legitimate communication and sensing.
 Furthermore, MA positioning also enables fine-grained control of channel power distributions. As shown in Figs.~\ref{MA_Mechanism}(d) and \ref{MA_Mechanism}(e), by moving antenna elements toward favorable propagation locations, the legitimate-link channel gain increases from $\text{-62.6}$ dB to $\text{-58.4}$ dB, whereas the eavesdropping-link gain decreases from $\text{-70.6}$ dB to $\text{-72.2}$ dB. These results demonstrate that MAs improve secrecy by jointly reshaping channel strength and spatial correlation.

\vspace{-3.5mm}
\subsection{Spatial Gains via Multi-Static Cooperation}

The performance gain of multi-static cooperation originates not merely from adding more nodes, but from exploiting geographically distributed propagation and observation perspectives. Spatially separated BSs observe legitimate users, sensing targets, and potential eavesdroppers through different propagation paths and spatial signatures, providing additional diversity and spatial DoFs for joint communication, sensing, and PLS design.
From the communication perspective, distributed transmission and reception enable cooperative beamforming and more flexible interference management. From the sensing perspective, multiple observation geometries improve target localization and parameter estimation. From the security perspective, distributed spatial diversity provides additional flexibility for secure beamforming and AN allocation, allowing legitimate links to be strengthened while unwanted signal leakage is suppressed from multiple directions.
MAs complement this macro-scale diversity by providing fine-grained control over the local propagation environment of each node. Multi-static cooperation determines the network-level transmitter-receiver geometry, whereas MA positioning reshapes local channel gains and correlations. Therefore, their coordination converts antenna mobility from a local adaptation mechanism into a network-level spatial-control capability for jointly improving communication, sensing, and PLS.

\vspace{-3.5mm}
\subsection{Mutually Reinforcing Closed-Loop of Communication, Sensing, and Security}

When eavesdropper locations or CSI are unavailable, conventional PLS designs often rely on conservative spatial protection, which can waste spatial resources. The proposed architecture instead exploits information exchange among communication, sensing, and security functions to support more adaptive physical-layer protection.
\begin{itemize}
\item \textbf{Sensing-Assisted Security:} 
When suspicious devices or active attackers generate observable echoes or interference signals, multi-static sensing can provide coarse location or angular information that reduces adversary uncertainty and assists security-oriented beamforming and AN design. However, for fully passive and silent eavesdroppers whose locations and CSI are unavailable, sensing alone cannot guarantee accurate localization. In such cases, robust beamforming and AN strategies based on statistical CSI or uncertainty-aware channel models remain necessary.
\item \textbf{Sensing-Assisted Communication:} 
In high-mobility scenarios, sensing information about user and target dynamics can assist predictive beam tracking and resource adaptation, reducing beam alignment latency and training overhead, as well as improving the reliability and continuity of secure transmission.
\item \textbf{Security-Assisted Sensing and Communication:} 
PLS mechanisms suppress information leakage and wireless interference during the exchange of sensing observations and coordination information, thereby improving the reliability of communication and sensing operations.
\end{itemize}
These interactions form a physical-layer closed loop in which sensing provides partial spatial information, while MA-enabled channel reconfiguration and multi-static diversity exploit it to improve confidentiality and sensing reliability. Overall, the proposed framework realizes dual-scale spatial reconfiguration: MAs reshape local propagation channels, multi-static cooperation provides network-level spatial diversity, and cross-functional information exchange adaptively exploits the resulting spatial DoFs for communication, sensing, and PLS.

\vspace{-2.5mm}
\section{Key Enabling Technologies}
This section highlights the key enabling technologies required to translate the proposed dual-scale spatial reconfiguration framework into practical systems.

\vspace{-3.5mm}
\subsection{Joint Design of Communication Beamforming, AN, and MA Positions}
In multi-static cooperative secure ISAC systems, MA positioning, communication beamforming, and AN transmission are strongly coupled. MA positions reshape the effective propagation channels, while beamforming and AN exploit the resulting spatial characteristics to strengthen legitimate transmission and suppress information leakage. Their joint design is therefore essential, but the resulting problem is highly non-convex because of continuous antenna-position variables, distributed cooperation, and coupled communication, sensing, and security requirements. Different solution paradigms are thus appropriate for different network dynamics:
\begin{itemize}
    \item \textbf{Alternating Optimization (AO) for Static Scenarios:}
    For slowly varying networks, centralized AO can iteratively update MA positions, beamforming and AN variables. With MA positions fixed, beamforming and AN variables can be optimized; then, MA positions can be updated based on the resulting channel conditions. Although such iterative approaches can achieve effective solutions, their computational latency limits rapid adaptation in highly dynamic environments.
    \item \textbf{Artificial Intelligence (AI)-Driven Methods for Dynamic Networks:} In fast time-varying scenarios, such as V2X and low-altitude UAV networks, the latency of iterative optimization may be too high to track rapidly changing network states. Learning-based controllers, including deep reinforcement learning for MA-aided secure ISAC~\cite{11020999}, can learn mappings from observed network states to beamforming, AN, and MA-control actions, thereby reducing online decision latency. For large-scale multi-static networks, multi-agent learning can further distribute decision-making among cooperating nodes. Practical deployment, however, must account for imperfect observations, environmental changes, and model generalization to ensure reliable spatial-control decisions.
\end{itemize}

\begin{table*}[t]
\centering
\caption{Comparison of Three Data Fusion Levels in Multi-Static Cooperative ISAC}
\label{data_fusion}
\footnotesize
\renewcommand{\arraystretch}{1.1}
\setlength{\tabcolsep}{2pt}

\begin{tabular}{
@{}|
>{\centering\arraybackslash}m{3.15cm}|
>{\centering\arraybackslash}m{4.780cm}|
>{\centering\arraybackslash}m{4.780cm}|
>{\centering\arraybackslash}m{4.780cm}|
@{}}

\hline
\textbf{Design Aspect} & \textbf{Signal-Level Fusion} & \textbf{Feature-Level Fusion} & \textbf{Decision-Level Fusion}
\\ \hline

\textbf{Fusion Object} 
& Raw echo samples or demodulated symbols & Estimated parameters or feature vectors & Local decisions, tracks, or confidence values
\\ \hline
\textbf{Information Retention}
&Highest & Medium & Lowest
\\ \hline

\textbf{Synchronization Requirement}
& Tight time, frequency, sampling, and phase synchronization & Frame-level synchronization and feature alignment & Loose coordination at decision or track level
\\ \hline

\textbf{Fronthaul Traffic}
& High & Moderate & Low
\\ \hline

\textbf{Processing Burden}
& High & Moderate & Low
\\ \hline

\textbf{Sensing Performance}
& Highest & Medium & Lower
\\ \hline

\textbf{Typical Deployment}
& Small tightly coordinated terrestrial clusters & Medium-scale terrestrial and air-ground cooperative networks & Wide-area or backhaul-limited heterogeneous networks
\\ \hline

\end{tabular}
\vspace{-4.75mm}
\end{table*}

\vspace{-6mm}
\subsection{Multi-Timescale Coordination and Control}

In practical MA-enabled multi-static secure ISAC systems, network configuration, MA positioning, and electronic transmission control are tightly coupled but operate at different rates. A multi-timescale coordination framework is therefore needed to balance spatial controllability, security performance, and implementation overhead.
\begin{itemize}
    \item \textbf{Slow-Timescale Network Planning:}
    Network-level decisions, such as cooperating-node selection, sensing-role assignment, and fusion-level configuration, mainly depend on long-term conditions including coverage requirements, hardware availability, fronthaul capacity, and security demands. These configurations can therefore be updated only when network topology, service requirements, or resource availability change significantly.
    \item \textbf{Medium-Timescale MA Reconfiguration:}
    MA positioning introduces additional spatial controllability but also incurs mechanical movement, calibration, and energy costs. Hence, antenna positions should respond to persistent changes in user geometry, propagation conditions, or security requirements rather than instantaneous channel fluctuations, avoiding excessive repositioning while retaining channel-reconfiguration gains.
    \item \textbf{Fast-Timescale Electronic Adaptation:}
    With MA positions fixed over short intervals, beamforming, AN allocation, waveform parameters, and scheduling can be updated rapidly according to instantaneous CSI and sensing requirements. Electronic adaptation handles fast channel variations, while mechanical MA reconfiguration is reserved for slower spatial changes.
\end{itemize}
Such multi-timescale coordination separates slow network-level configuration, medium-timescale spatial reconfiguration, and fast electronic adaptation, thereby reducing unnecessary MA movement, calibration, and signaling overhead. It also provides a practical control architecture for heterogeneous multi-static networks with different mobility levels, fronthaul capabilities, and service requirements.

\vspace{-3mm}
\subsection{Channel State Information Acquisition}

Accurate channel information is essential for MA positioning, cooperative beamforming, and sensing processing, but its availability differs across links. For legitimate users, pilot-assisted sounding can be performed at selected anchor positions within the MA movement region, while sensing-related propagation characteristics can be inferred from echoes of known ISAC waveforms. To avoid exhaustive spatial sounding, compressed sensing or learning-assisted electromagnetic-field reconstruction can estimate channel variations over the movement region from limited measurements.
Passive eavesdroppers present a fundamentally different challenge because they transmit no pilots and may deliberately conceal their presence; hence, their instantaneous CSI is generally unavailable. Depending on adversary observability, the system may rely on estimated locations, statistical CSI, or bounded uncertainty regions. When adversaries generate observable reflections or interference signals, multi-static sensing can provide partial spatial information to assist security-oriented beamforming and AN design. With incomplete CSI, MA positioning, beamforming, and AN allocation should instead be optimized robustly over the corresponding uncertainty region. For completely unknown passive eavesdroppers, eavesdropper-specific nulling is infeasible, and MA-induced spatial controllability together with multi-static diversity should be exploited to improve secrecy robustness.
Practical CSI acquisition must also account for MA positioning and channel reconstruction errors, which can cause channel mismatch and phase deviations, especially at millimeter-wave and terahertz frequencies. Robust designs should incorporate these residual uncertainties to maintain reliable communication, sensing, and PLS performance.

\vspace{-4mm}
\subsection{Network Synchronization and Multi-Level Data Fusion}

Multi-static secure ISAC requires accurate spatiotemporal coordination among distributed nodes, creating a practical trade-off between synchronization accuracy and coordination overhead~\cite{10684491}. Clock offsets among BSs introduce timing and phase mismatches that degrade coherent beamforming and sensing processing, while MA positioning errors can further cause non-negligible phase deviations, particularly at millimeter-wave and terahertz frequencies. Therefore, practical deployments require scalable synchronization and robust system designs that account for timing, phase, and position uncertainties to maintain reliable communication, sensing, and PLS performance.
Multi-static cooperation also requires efficient exchange and fusion of sensing information among distributed nodes. Depending on fronthaul capacity, synchronization capability, computational resources, and sensing requirements, three hierarchical fusion levels can be adopted, as summarized in Table~\ref{data_fusion}.
\begin{itemize}
\item \textbf{Signal-Level Fusion:}
The fusion objects are raw echo samples or demodulated symbols from distributed nodes. By preserving the most complete sensing observations, signal-level fusion offers the highest potential for coherent processing gain and high-resolution sensing~\cite{11203895}. However, it requires tight time, frequency, sampling, and phase synchronization, together with high-capacity fronthaul for transporting and processing raw data. Therefore, it is mainly suitable for tightly coordinated cooperation clusters with sufficient communication and computational resources, where the sensing accuracy requirement justifies the associated overhead.
\item \textbf{Feature-Level Fusion:}
Each BS first performs local sensing processing to extract intermediate features, such as target range, angle, velocity, or radar cross section, and then exchanges these compact representations for fusion~\cite{11442912}. Compared with signal-level fusion, feature-level fusion significantly reduces fronthaul traffic and relaxes the requirement for strict carrier-phase synchronization, while still requiring reliable local estimation and feature consistency among cooperating nodes. It therefore provides a practical balance between sensing accuracy, communication overhead, and implementation complexity, making it suitable for medium-scale terrestrial and air-ground cooperative deployments.
\item \textbf{Decision-Level Fusion:}
When fronthaul or synchronization resources are severely constrained, each node can independently perform local detection or tracking and exchange only compact decision outputs, such as target-presence indicators, tracks, or confidence values~\cite{10666707}. This mode minimizes data exchange and is less sensitive to synchronization errors, improving scalability and robustness in heterogeneous networks. However, the loss of detailed sensing information limits the achievable fusion gain, and the final performance depends strongly on local sensing reliability and the adopted decision strategy. It is therefore more suitable for wide-area or resource-constrained deployments where sensing availability and system resilience are prioritized.   
\end{itemize}
The fusion level should be selected according to the operating conditions of the cooperative network. Signal-level fusion is preferred when accurate synchronization and sufficient fronthaul and computing resources are available, feature-level fusion provides a favorable compromise under moderate constraints, and decision-level fusion becomes preferable when communication or coordination resources are limited. In dynamic networks, persistent changes in synchronization quality, fronthaul availability, computational capacity, or sensing requirements may trigger a transition among these operating modes. Such adaptation should also account for the signaling, processing, and coordination overhead introduced by fusion reconfiguration, thereby avoiding unnecessary switching and maintaining stable operation.

\begin{figure}[t!]
	\centering
	\includegraphics[height=4.55cm]{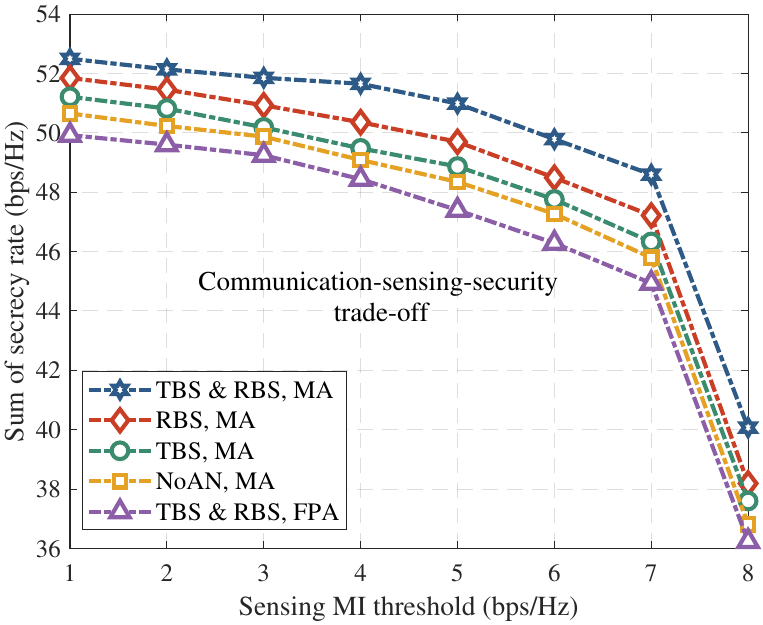}
	\vspace{-0.85em}
	\caption{SSR versus sensing MI threshold.}
    \label{SSR_MI}
    \vspace{-5mm}
\end{figure}

\vspace{-4mm}
\section{Case Study and Engineering Insights}

To demonstrate the engineering benefits of the proposed framework, we consider a quasi-static 3D multi-static ISAC network with two transmitting BSs (TBSs) and two receiving BSs (RBSs). The scenario includes two downlink users, two uplink users, one sensing target, two communication eavesdroppers, two sensing eavesdroppers, and two clutter scatterers. The TBSs are located at $(0,50,20)$ m and $(50,100,20)$ m, the RBSs at $(50,0,20)$ m and $(100,50,20)$ m, and the sensing target at $(50,50,20)$ m, while the remaining nodes are randomly distributed within the coverage area. The carrier frequency is 3 GHz, and each BS employs $6$ MAs within a $4\lambda\times4\lambda$ movement region with a minimum spacing of $\lambda/2$. The maximum downlink and uplink transmit powers are 40 dBm and 10 dBm, respectively, and the noise power is $-90$ dBm. All wireless links follow Rician fading. To isolate the potential gains of spatial reconfiguration and multi-static cooperation, perfect CSI at the CP and ideal synchronization among the cooperating BSs are assumed, providing a performance upper bound for the proposed MA-enabled multi-static ISAC system. The sum of secrecy rates (SSR) is maximized by jointly optimizing receive combiners, transmit and AN beamforming, uplink power allocation, and MA positions, subject to transmit power, sensing mutual information (MI), target detection probability at the sensing eavesdroppers, MA movement region, and minimum spacing constraints.

\begin{figure}[t!]
	\centering
	\includegraphics[height=4.55cm]{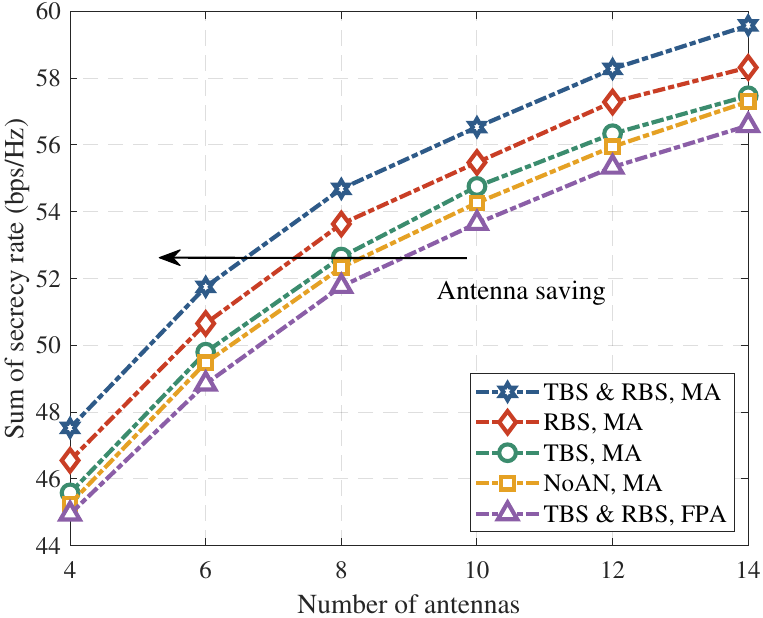}
	\vspace{-0.85em}
	\caption{SSR versus number of antennas.}
    \label{SSR_N}
    \vspace{-5mm}
\end{figure}

We compare five representative configurations: FPA at both the TBS and RBS sides, MA at the TBS only, MA at the RBS only, MA at both sides without AN, and the full MA configuration at both sides with AN.
Fig.~\ref{SSR_MI} illustrates the communication-sensing-security trade-off under different sensing requirements. When the sensing MI requirement is relatively relaxed, all configurations maintain high SSR values. As the sensing constraint becomes more stringent, the FPA-based configuration exhibits a more pronounced degradation due to its limited spatial controllability, whereas MA-enabled schemes maintain better secrecy performance by exploiting MA-induced channel reconfiguration. Specifically, TBS-side MAs provide additional flexibility to reshape the transmitted spatial distribution and reduce potential information leakage toward eavesdroppers, while RBS-side MAs improve legitimate signal reception by exploiting favorable propagation conditions. The full MA configuration achieves the best overall SSR by jointly exploiting spatial controllability at both sides, and its advantage over the MA configuration without AN further demonstrates the complementary role of AN in enhancing physical-layer protection.
Fig.~\ref{SSR_N} further investigates the hardware efficiency. To achieve an SSR of approximately 52 bps/Hz, the FPA configuration requires about 8 antennas, whereas the full MA configuration achieves comparable performance with about 6 antennas, corresponding to an approximately 25\% reduction in antenna elements. 
The result indicates that MA-based spatial reconfiguration can alleviate the need for continuously increasing antenna numbers by providing additional spatial controllability through antenna-position adaptation. By exploiting position-induced channel variations, MAs improve spatial discrimination between legitimate and eavesdropping links without relying solely on additional antenna elements. Therefore, MA technology provides a promising approach for reducing antenna scaling requirements in future secure ISAC deployments.

\section{Open Challenges and Future Directions}
Despite the potential of MA-enabled multi-static cooperative secure ISAC, several challenges remain before practical deployment. This section highlights four key directions for future research.

\vspace{-6mm}
\subsection{Near-Field Effects and Range-Domain Security}

As 6G systems move toward higher frequency bands, larger apertures, and shorter communication and sensing ranges, near-field spherical-wave propagation will become increasingly relevant. Unlike far-field designs that mainly exploit angular separation, near-field propagation introduces additional range-domain spatial DoFs, potentially enabling legitimate users and eavesdroppers with similar angular directions but different ranges to be spatially separated. Exploiting this capability requires near-field channel models, beamforming strategies, and security metrics that jointly account for communication, sensing, and confidentiality. Extending MA-enabled multi-static secure ISAC to near-field propagation represents an important research direction.

\vspace{-5mm}
\subsection{Scalable Coordination and Practical Implementation}

Large-scale multi-static deployments require scalable coordination under increasing network size, heterogeneous fronthaul/backhaul conditions, and diverse service requirements. Fully centralized coordination can be effective for small, tightly coordinated clusters, but its signaling, computation, and synchronization overhead increase with network scale. Hierarchical or distributed architectures are therefore needed, where local BSs or clusters perform beamforming, sensing processing, and MA reconfiguration using locally available information, while network-level coordination manages cooperating-node selection, resource allocation, and inter-cluster information exchange. Practical MA deployment also introduces hardware-specific constraints, including position calibration, mechanical actuation, energy consumption, and finite repositioning latency. These constraints limit the feasible MA reconfiguration rate and may reduce the gains predicted under idealized models. Future research should therefore jointly account for coordination scalability, hardware constraints, and spatial-reconfiguration gains in practical system design.

\vspace{-4mm}
\subsection{Dynamic Environment and Robust Secure Control}

Practical networks operate in dynamic environments where users, sensing targets, and adversaries may continuously change their locations, altering the spatial relationships among communication, sensing, and security links. Meanwhile, finite MA response time prevents instantaneous adaptation to such variations. Robust and predictive control mechanisms are therefore needed to jointly coordinate MA reconfiguration, beamforming, and AN under mobility and channel uncertainty. A key challenge is to maintain physical-layer secrecy and sensing reliability during the transition period between environmental variations and spatial reconfiguration, where the available spatial DoFs may not yet be fully exploited.

\vspace{-4mm}
\subsection{Cross-Layer Security Considerations}

This article primarily focuses on PLS against wireless eavesdropping and jamming. Practical multi-static ISAC networks also rely on distributed CSI sharing, sensing data exchange, and coordination signaling, which introduce additional security requirements beyond physical-layer protection. PLS cannot replace authentication, integrity protection, access control, or secure cooperative signaling. Moreover, AI-assisted control may be affected by inaccurate or manipulated sensing information, CSI inputs, and unreliable model outputs, leading to inappropriate beamforming, AN, or MA-control decisions. Future research should therefore investigate cross-layer frameworks that combine PLS with trustworthy information exchange, secure control signaling, and robust AI-assisted decision making without sacrificing the spatial flexibility provided by MAs and multi-static cooperation.

\section{Conclusion}

This article has presented an MA-enabled multi-static cooperative secure ISAC framework centered on dual-scale spatial reconfiguration. By coupling micro-scale MA-induced channel reconfiguration with macro-scale multi-static spatial diversity, the framework establishes a dual-scale spatial control capability for jointly enhancing communication, sensing, and PLS. We have elucidated the underlying physical-layer gain mechanisms and discussed the key enabling technologies for practical implementation. The case study further demonstrates that MA-enabled spatial reconfiguration can improve the communication-sensing-security trade-off and reduce the required antenna resources compared with conventional FPAs. Finally, we have outlined open challenges and future directions, envisioning that the deep integration of MA and multi-static cooperative ISAC will play a pivotal role in resilient and secure 6G networks.

\balance
\bibliographystyle{IEEEtran}


\end{document}

%% file: SupportDocuments/YFPreamble.tex
\usepackage{amssymb}
\usepackage{amsmath}
\usepackage{overpic}
\usepackage{algorithm}
\usepackage{algorithmic}
\usepackage{multirow}
\usepackage{array}
\usepackage{graphicx}
\usepackage{bm}
\usepackage{color}
\usepackage{tikz}
\usepackage{epstopdf}
\usepackage{stfloats}
\usepackage{cite}
\usepackage{tikz}
\usepackage{amsmath}
\usepackage{psfrag}
\usepackage{acronym}
\usepackage{enumerate}
\usetikzlibrary{arrows}
\usetikzlibrary{decorations}

\definecolor{BLUE}{rgb}{0,0,1}

\acrodef{aoa}[AOA]{angle-of-arrival}
\acrodef{bcrb}[BCRB]{Bayesian Cram\'{e}r-Rao bound}
\acrodef{bfim}[BFIM]{Bayesian Fisher information matrix}
\acrodef{bp}[BP]{belief propagation}
\acrodef{cdi}[CDI]{cooperative dilution intensity}
\acrodef{cir}[CIR]{channel impulse response}
\acrodef{cl}[CL]{cooperative localization}
\acrodef{cp}[CP]{cyclic prefix}
\acrodef{crb}[CRB]{Cram\'{e}r-Rao bound}
\acrodef{crlb}[CRLB]{Cram\'{e}r-Rao lower bound}
\acrodef{dft}[DFT]{discrete Fourier transform}
\acrodef{dof}[DoF]{degree of freedom}
\acrodef{dpeb}[DPEB]{directional position error bound}
\acrodef{fim}[FIM]{Fisher information matrix}
\acrodef{efim}[EFIM]{equivalent Fisher information matrix}
\acrodef{hogs}[HOGS]{hybrid orthogonal-Gaussian signalling}
\acrodef{hsug}[HSUG]{hybrid semi-unitary--Gaussian}
\acrodef{ici}[ICI]{information coupling intensity}
\acrodef{icrb}[ICRB]{inverse CRB}
\acrodef{iid}[i.i.d.]{independently and identically distributed}
\acrodef{im}[IM]{index modulation}
\acrodef{isac}[ISAC]{integrated sensing and communication}
\acrodef{los}[LoS]{line-of-sight}
\acrodef{mse}[MSE]{mean-squared error}
\acrodef{ofdm}[OFDM]{orthogonal frequency-division multiplexing}
\acrodef{pdf}[PDF]{probability density function}
\acrodef{peb}[PEB]{position error bound}
\acrodef{speb}[SPEB]{squared position error bound}
\acrodef{pll}[PLL]{phase-locked loop}
\acrodef{psk}[PSK]{phase shift keying}
\acrodef{p2p}[P2P]{point-to-point}
\acrodef{qam}[QAM]{quadrature amplitude modulation}
\acrodef{rbs}[RBS]{reference broadcast synchronization}
\acrodef{rhs}[RHS]{right hand side}
\acrodef{rii}[RII]{ranging information intensity}
\acrodef{rss}[RSS]{received signal strength}
\acrodef{rc}[RC]{ranging coefficient}
\acrodef{speb}[SPEB]{squared position error bound}
\acrodef{toa}[TOA]{time-of-arrival}
\acrodef{tdoa}[TDOA]{time-difference-of-arrival}
\acrodef{tpsn}[TPSN]{time synchronization protocol for sensor network}
\acrodef{vmp}[VMP]{variational message passing}
\acrodef{wsn}[WSN]{wireless sensor network}
\acrodef{efim}[EFIM]{equivalent Fisher information matrix}
\acrodef{dio}[DIO]{distance-information-only}
\acrodef{aio}[AIO]{angle-information-only}
\acrodef{saaf}[SAAF]{squared array aperture function}
\acrodef{snc}[S\&C]{sensing and communications}
\acrodef{uoa}[UOA]{uniformly oriented array}
\acrodef{rgg}[RGG]{random geometric graph}
\acrodef{snr}[SNR]{signal-to-noise ratio}
\acrodef{eoc}[EoC]{efficiency of cooperation}
\acrodef{npi}[NPI]{nominal position information}
\acrodef{gnss}[GNSS]{global navigation satellite system}
\acrodef{mimo}[MIMO]{multiple-input multiple-output}
\acrodef{mcs}[MCS]{minimally constrained system}
\acrodef{zzb}[ZZB]{Ziv-Zakai lower bound}
\acrodef{wwb}[WWB]{Weiss-Weinstein lower bound}
\acrodef{nlos}[NLOS]{non-light-of-sight}
\acrodef{mmse}[MMSE]{minimum mean squared error}
\acrodef{uav}[UAV]{unmanned aerial vehicle}
\acrodef{ppp}[PPP]{Poisson point process}
\acrodef{bpp}[BPP]{binomial point process}
\acrodef{cln}[CLN]{cooperative location-aware network}
\acrodef{pdr}[PDR]{pedestrian dead reckoning}
\acrodef{ml}[ML]{maximum likelihood}
\acrodef{map}[MAP]{maximum \textit{a posteriori}}
\acrodef{kkt}[KKT]{Karush-Kuhn-Tucker}
\acrodef{st}[ST]{subspace tradeoff}
\acrodef{drt}[DRT]{deterministic-random tradeoff}
\acrodef{ustm}[USTM]{unitary space-time modulation}
\acrodef{pri}[PRI]{pulse repetition interval}
\acrodef{prf}[PRF]{pulse repetition frequency}
\acrodef{acf}[ACF]{autocorrelation function}
\acrodef{isl}[ISL]{integrated sidelobe level}
\acrodef{islr}[ISLR]{integrated sidelobe level ratio}
\acrodef{rgi}[MFI]{mainlobe fluctuation intensity}
\acrodef{irgi}[AMFI]{averaged mainlobe fluctuation intensity}
\acrodef{cds}[CDS]{cyclic difference set}
\acrodef{psl}[PSL]{peak sidelobe level}
\acrodef{aesl}[AESL]{average expected sidelobe level}
\acrodef{pesl}[PESL]{peak expected sidelobe level}
\acrodef{masm}[MASM]{MASked Modulation}
\acrodef{si}[SI]{self-interference}
\acrodef{cpi}[CPI]{coherent processing interval}

\acrodefplural{dof}[DoFs]{degrees of freedom}
\acrodefplural{drt}[DRTs]{deterministic-random tradeoffs}
\acrodefplural{cds}[CDSs]{cyclic difference sets}
